\documentclass[aps,a4paper,showpacs]{revtex4}
\usepackage{graphicx}
\usepackage{amsmath}
\usepackage{amssymb}
\usepackage{enumerate}
\usepackage{subfigure}
\usepackage{tabularx}
\usepackage{color}

\newcommand{\be}{\begin{equation}}
\newcommand{\ee}{\end{equation}}
\newcommand{\ben}{\begin{eqnarray}}
\newcommand{\een}{\end{eqnarray}}
\newcommand{\bes}{\begin{subequations}}
\newcommand{\ees}{\end{subequations}}

\newcommand{\bb}{\bibitem}

\begin{document}
\title{Topological Excitations in Magnetic Materials}
\author{D. Bazeia$^{1}$\footnote{Corresponding author. Email: bazeia@fisica.ufpb.br}, M.M. Doria$^{2,3}$ and E.I.B. Rodrigues$^1$}
\affiliation{$^1$Departamento de F\'\i sica, Universidade Federal da Para\'\i ba, 58051-970 Jo\~ao Pessoa, PB, Brazil}
\affiliation{$^2$Instituto de F\'\i sica, Universidade Federal do Rio de Janeiro, Rio de Janeiro, Brazil}
\affiliation{{$^3$Dipartimento di Fisica, Universit\`a di Camerino, I-62032 Camerino, Italy}}
\begin{abstract}
In this work we propose a new route to describe topological excitations in magnetic systems through a single real scalar field. We show here that spherically symmetric structures in two spatial dimensions, which map helical excitations in magnetic materials, admit this formulation and can be used to model skyrmion-like structures in magnetic materials.
\end{abstract}
\date{\today}
\maketitle


\section{Introduction}

Magnetic materials \cite{book} are rich enough to support complex localized topological structures known as skyrmions \cite{skyrme,ms}. Indeed, depending of the type of material, the localized excitations may appear as isolated structures or as spatially organized arrays or lattices; see, e.g, Refs.~\cite{s0,s1,s11,s2,s3,s4,s5,s6}, for theoretical investigations, and Refs.~\cite{e1,e2,e3,e4} for the experimental observations of skyrmions and lattices of skyrmions. See also \cite{nano1,nano2,size2,size,prl15,W} and references therein, for other interesting investigations on the subject.

We go back in time to the 1960's and note that, soon after the proposal of skyrmions, Hobard \cite{hob} and Derrick \cite{der} stablished with simple scaling arguments that localized particle-like structures are not stable when described by scalar fields in dimension greater then one.
To circumvent the scaling obstruction, the authors of \cite{laf} have suggested modification of the scalar field kinematics. More recently, in \cite{bmm} one proposes a different route, relying on taking the scalar field with standard kinematics, but changing the way the self-interactions enter the game, with explicit dependence on the spatial coordinates. With the proposal of \cite{bmm}, it was possible to suggest the presence of stable, spherically symmetric structures in arbitrary spatial dimension, constructed from a single real scalar field.

In the current work we follow \cite{bmm} to study the case of planar solutions which are skyrmion-like structures and can be used to model magnetic excitations in planar magnetic metals such as the ones studied in \cite{s11,s2,s3,s6,mn}. Here we focus on the presence of analytical solutions, paying attention to the case of excitations with the magnetization having constant modulus, and we postpone to a future work other possibilities, in particular the case where the modulus of the magnetization may vary.

We organize the work such that in Sec.~II we describe some basic facts about skyrmions, and adjust the degrees of freedom needed to model the planar skyrmion-like excitations which we suppose are present in the magnetic material. In Sec.~III we deal with models that describe the scalar field in two spatial dimensions, as suggested in \cite{bmm}. There, we introduce two distinct models and find analytical results for the corresponding field configurations and energy density. We also study stability of the solutions and show an interesting way to describe the magnetization in the case where it is a vector of constant modulus, capable of mapping the presence of helical excitations in magnetic materials. The two models have different solutions, and map magnetic excitations that connect distinct ground states, engendering distinct properties, which appear associated with the corresponding skyrmion numbers. We also illustrate how the helical profile appears in each case. We end the work in Sec.~IV, where we include our comments and conclusions.

\section{Generalities}

 Let us follow \cite{s2,s3,s6,mn} and comment on the planar model of interest in this work. In general, in models where the magnetization vector is homogeneous along the z direction, the cylindrically symmetric magnetization is described by the two main degrees of freedom, one being the local amplitude of the magnetic moments and the other, the direction in the plane, both as function $r$, the distance to the cylinder axis in cylindrical coordinates. The standard procedure then relies on writing a free energy density, usually based on Dzyaloshinsky-Moriya interactions \cite{dm,dm2} to obtain the equations of motion for the main degrees of freedom. This is a line of investigation, and that leads to two coupled second-order nonlinear differential equations. The specific form of the equations depend on the model under consideration, and usually require numerical investigations. The models engenders several parameters, which are in general used to control the stiffness, strength and other physical properties of the magnetic interactions in the magnetic material.

In the current work we want to offer an alternative description to the numerical investigation, focusing mainly on the construction of analytical results. As usual, we consider that the magnetization is homogeneous along the ${\hat{z}}$ direction, but we also suppose that it has constant modulus. Thus, the normalized magnetization vector has the form ${\bf M}=M_1 \hat{{x}}+ M_2\hat{{y}}+M_3\hat{{z}}$, and should obey the constraint $M_1^2(x,y)+M_2^2(x,y)+M_3^2(x,y)=1$, which reduces the problem to two degrees of freedom. 

We will further reduce the problem to a single degree of freedom, supposing that the magnetization ${\bf M}$ is a unit vector ortogonal to the radial direction ${\hat{r}}$, such that the model now engenders helical profile, which is known to be present in systems with $D_{2d}$ symmetry, with the behavior reminding us of the magnetization vector for Bloch walls in a diversity of ferromagnetic materials \cite{book,s2}. The helical profile which we investigate in the current work is inspired from \cite{s2}, where it is well illustrated. This type of structure appears, for instance, in magnetic metals such as MnSi; see, e.g., \cite{s3,mn}. In cylindrical coordinates, the model is then described by ${\bf M}=M_1(r,\theta){\hat z}+ M_2(r,\theta){\hat\theta}$, with $M_1^2(r,\theta)+M_2^2(r,\theta)=1$. 

However, instead of describing the problem with two degrees of freedom and a constraint, we will model it via the dynamical scalar field that appears in \cite{bmm}, in the case of two spatial dimensions. Since we are dealing with helical excitations, we have to consider the magnetization ${\bf M}$ as a unit vector orthogonal to the radial direction. We use cylindrical coordinates to model this situation with the magnetization given by
\be\label{mm}
{\bf M}={\hat z}\sin\left(\frac{\pi}{2}\phi+\delta\right)+{\hat\theta} \cos\left(\frac{\pi}{2}\phi+\delta\right),
\ee
where $\phi$ is a scalar field that may vary in the $(r,\theta)$ plane, and $\delta$ is a constant. If $\phi$ is a uniform solution $\phi=constant$, we have the magnetization vector pointing in a fixed direction throughout the magnetic material, and this represents a ground state. The constant phase $\delta$ can be used to adequately describe the ground state of the system. Thus, we have to search for field configurations that vary in space, leading to the construction of magnetic solutions, that connect distinct ground states and have non vanishing energy density and positive energy, when compared to the energy of the ground states.

Moreover, since we are searching for skyrmions, we follow the usual route, and we introduce the skyrmion number, the topological quantity that usually appears in the investigation. It is given by
\be\label{sky}
Q=\frac{1}{8\pi}\int d^2{x} \;\;{\bf M}\cdot{\partial_i{\bf M}}\times {\partial_j{\bf M}}\;\varepsilon^{ij},
\ee
where $i,j=1,2$, and $\partial_i$ stands for the derivative with respect to $x^i$, with $\epsilon^{ij}$ being the completely antisymmetric tensor such that $\epsilon^{12}=1$. This topological quantity will guide us to the search of skyrmions in the models that we investigate below.

Before investigating the field configuration, however, let us comment a little further on the behavior of the magnetic solutions that we are going to study below. We see from Eqs.~\eqref{mm} and \eqref{sky} that the scalar field determines both the magnetization and the skyrmion number. Moreover, the energy associated to the field configuration depends on the specific model and the basic symmetries which we use to describe it.  We then focus on this, and we model the magnetization with the scalar field investigated in \cite{bmm}, as we explain in details in the next Section.

\section{The models}
\label{secIII}

Let us now pay attention to the scalar field $\phi$ which we introduced in Eq.~\eqref{mm}. We suppose that it is described by the planar model of Ref.~\cite{bmm}. Thus, the Lagrange density ${\cal L}$ has the form
\be\label{model}
{\cal L}=\frac12\dot\phi^2-\frac12 \nabla\phi\cdot\nabla\phi-U(\phi),
\ee
where dot stands for time derivative, and $\nabla$ represents the gradient in the $(x,y)$ or $(r,\theta)$ plane. In the case of time independent and spherically symmetric configuration, $\phi=\phi(r)$, it is supposed to obey the equation of motion
\be
{\frac1r\frac{d}{dr}\left(r\frac{d\phi}{dr}\right)}=\frac{dU}{d\phi},
\ee
where $U=U(r,\phi)$ is given by
\be
U(r,\phi)=\frac1{2r^2}P(\phi),
\ee
with $P(\phi)$ an even polynomial which contains nongradient terms in $\phi$. 
The energy for solution $\phi=\phi(r)$ is given by
\be
E= 2\pi \int_0^\infty r dr\,\rho(r) ,
\ee
with $\rho(r)$ being the energy density, such that
\be\label{eden}
\rho(r) =\frac12 \left(\frac{d\phi}{dr}\right)^2+ \frac{1}{2r^2}P(\phi).
\ee 
This is the framework we use to describe the scalar field $\phi$ that controls the magnetization which we proposed in \eqref{mm}.

We go on and get inspiration from the Landau theory to expand the nongradient contributions that appears from $P(\phi)$ to consider two different possibilities, which lead to two distinct models, described by
\be\label{p4}
P_4(\phi)=\frac{1}{(1-s)^2}({\bar s}^2-\phi^2)^2,
\ee
and
\be\label{p6}
P_6(\phi)=\frac{1}{(1-s)^2}\phi^2({\bar s}^2-\phi^2)^2,
\ee
where the dimensionless parameters $s\in[0,1)$ and ${\bar s}>0$ are introduced to control the strength of the interactions, which is important to make contact with realistic models. The two models engender spontaneous symmetry breaking, as it is required for the presence of topological structures: the first model presents interactions modulated by the fourth-order power in the field, and support two asymmetric ground states, $\bar{\phi}_\pm=\pm{\bar s}$. We will use these two ground states and the magnetization \eqref{mm} to guide us to describe magnetic excitations in this model in Sec.~\ref{planar}. The second model is different, and presents interactions modulated by the fourth- and sixth-order power, supporting a symmetric ground state, $\bar{\phi}_0=0$, and two asymmetric ones, $\bar{\phi}_\pm=\pm{\bar s}$. Also, we will use these ground states and the magnetization \eqref{mm} to guide us to describe magnetic excitations in this model, in Sec.~\ref{planar}. For simplicity, we will take ${\bar s}=1$ from now on.

We follow Ref.~\cite{bmm} and consider that the field, coordinates and parameters are all dimensionless, so we shall need to introduce a quantity to measure distance. This is an important issue, which we further deal with below. The models \eqref{p4} and \eqref{p6} are known to support kinklike solutions in one spatial dimension, and here we will show that they support stable radial configurations having energy density localized around a given radial position, depending on the value of $s$. We then associate $s$ with the characteristic size of the structure. We do this denoting the size ${\bar r}$ as the mean matter radius of the field configuration, defined as the radial distance $r$ weighted by the energy density of the static scalar field which we will use below to describe the skyrmion configuration. This mean matter radius is given by
\be\label{size}
{\bar r}=\frac{\int_0^\infty \rho(r) r^2 dr}{\int_0^\infty \rho(r)r dr}.
\ee
This definition is very natural, since we are working with static field and so the energy corresponds to the mass of the configuration. We can use it to describe the size of the skyrmion; see, e.g., Refs.~\cite{size2,size}, for other investigations concerning this issue.

\begin{figure}[t!]
\centerline{{\includegraphics[scale=0.8]{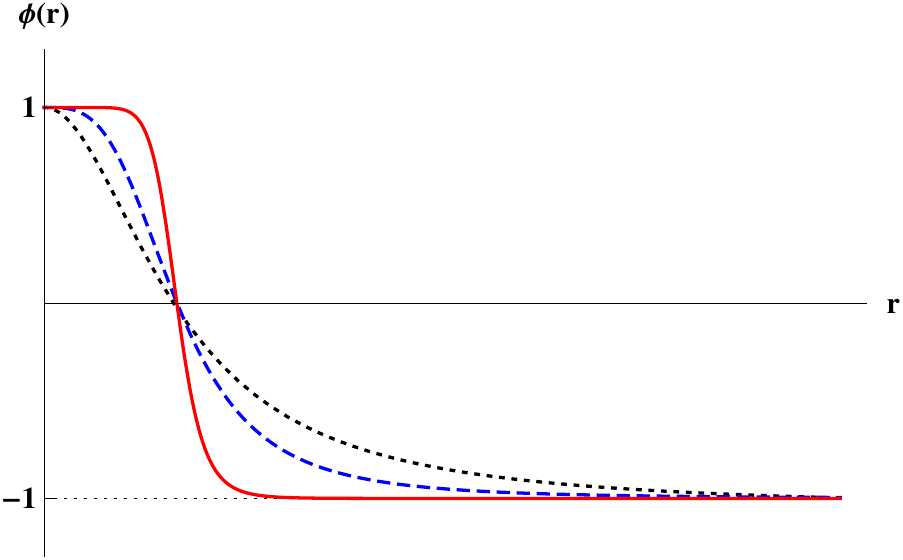}}\qquad\qquad\qquad{\includegraphics[scale=0.8]{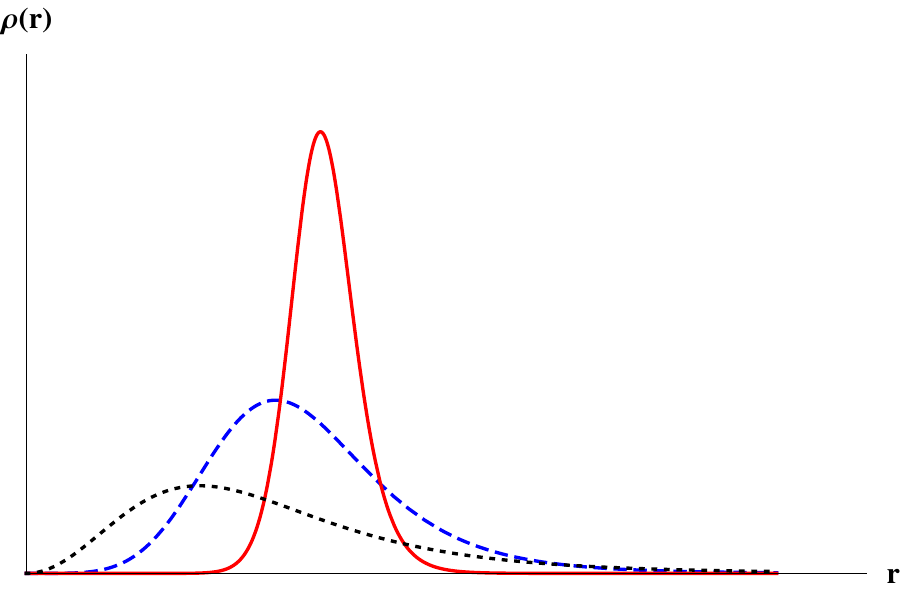}}}
\caption{(Color online) The solution (left panel) and the energy density (right panel) for the model \eqref{p4}, depicted for $s=0$ with black, dotted line, for $s=0.4$ with blue, dashed line, and for $s=0.8$ with red, solid line.}\label{fig1}
\end{figure}

We can also attain to the structure, a topological property. Since the field configuration at the origin $\phi(0)$ should represent a ground state, and since it should connect another ground state asymptotically, we must have $\phi(0)\neq\phi(\infty)$. We can describe the correct profile using specific boundary conditions, which will be of good use to define the skyrmion number associated to the magnetization vector, as we further investigate below. To explore this route, let us first deal with the model \eqref{p4}; we can write its equation of motion as
\be\label{em4}
r^2\frac{d^2\phi}{dr^2}+ r \frac{d\phi}{dr}+\frac{2\phi(1-\phi^2)}{(1-s)^2}=0.
\ee
We see that if $\phi$ is a solution, so is $-\phi$. We shall then concentrate on one of them. We consider the boundary conditions such that $\phi(0)=1$ and $\phi(\infty)=-1$, which are compatible with the form of the polynomial interactions that the model engender, as we see from \eqref{p4}, and from the equation of motion \eqref{em4} as well. We have been able to solve \eqref{em4} analytically for $s$ arbitrary, and the solution can be expressed in the form
\be\label{phi4}
\phi_{s}(r)=\frac{1-r^{2/(1-s)}}{1+r^{2/(1-s)}}.
\ee
As we commented before, there is another solution, with the minus sign, which behaves similarly. We depict the solution \eqref{phi4} in Fig.~\ref{fig1}, for some values of $s$.

The corresponding energy density $\rho(r)$ is given by \eqref{eden}, associated to the field theory model \eqref{model}. It has the form
\be
\rho_{s}(r)=\frac{16r^{2(1+s)/(1-s)} }{ (1-s)^2(1+r^{2/(1-s)})^4}.
\ee
We also depict the energy density in Fig.~\ref{fig1}. We see that it becomes sharper and sharper, as $s$ increases in the interval $[0,1)$. The total energy is given by
\be\label{ene4}
E=\frac{8\pi}{3(1-s)}.
\ee
It increases as $s$ increases in the interval $[0,1)$, and we note that the sharper the solution is, the higher the energy becomes.
Also, we follow \eqref{size} and introduce the size of the solution. It is ${\bar r}_s$, and it is given by
\be
{\bar r}_s=\frac{\pi(3-s)(1-s^2)}{8\cos\left(\frac{\pi}{2}s\right)}.
\ee

\begin{figure}[t]
\centerline{{\includegraphics[scale=0.8]{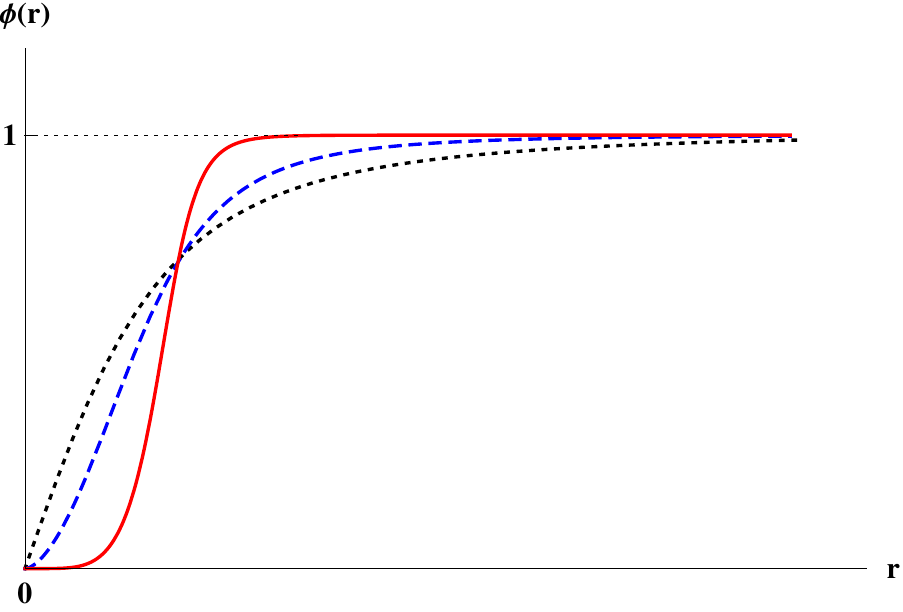}}\qquad\qquad\qquad{\includegraphics[scale=0.8]{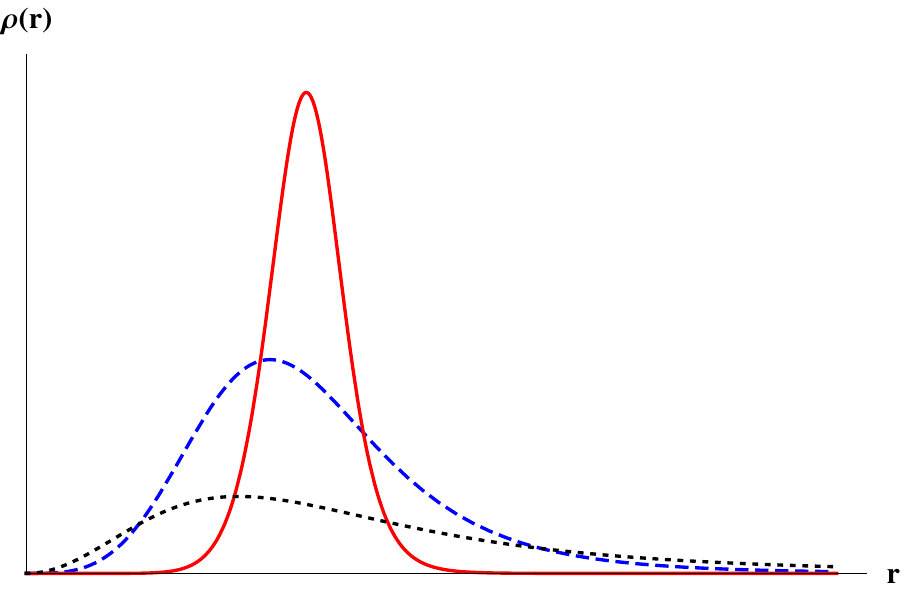}}}
\caption{(Color online) The solution (left panel) and the energy density (right panel) for the model \eqref{p6}, depicted for $s=0$ with black, dotted line, for $s=0.4$ with blue, dashed line, and for $s=0.8$ with red, solid line.}\label{fig2}
\end{figure}

For the second model \eqref{p6}, the equation of motion is
\be
r^2\frac{d^2\phi}{dr^2}+ r \frac{d\phi}{dr}-\frac{\phi(1-\phi^2)^2}{(1-s)^2}+\frac{2\phi^3(1-\phi^2)}{(1-s)^2}=0.
\ee
The situation here is different, since the model \eqref{p6} has three minima, one at zero, and two at $\pm1$. They are uniform solutions, which correspond to distinct ground state possibilities. Also, we note that if $\phi$ is solution, so is $-\phi$. Thus, we now search for solution that obey the boundary conditions $\phi(0)=0$ and $\phi(\infty)=1$. Such solution corresponds to a field configuration that connects the two ground states $\phi=0$ and $\phi=1$. We have been able to solve the equation of motion exactly, and we could write the solution in the form
\be\label{phi6}
\phi_s(r)={\frac{r^{1/(1-s)}}{\sqrt{1+r^{2/(1-s)}}}}.
\ee
We depict this solution in Fig.~\ref{fig2}, for some specific values of $s$.

The energy density is this case has the form
\be
\rho_s(r)= \frac{ r^{2/(1-s)}}{(1-s)^2(1+r^{2/(1-s)})^{3}},
\ee
which is also depicted in Fig.~\ref{fig2}. The total energy is given by
\be\label{ene6}
E=\frac{\pi^2 s}{2\sin(\pi s)}.
\ee
We see that it increases as $s$ increases in the interval $[0,1)$, and we note that the sharper the solution is, the higher the energy becomes; see Fig.~\ref{fig2}. The behavior is similar to the case observed in the previous model. In this case, the mean matter radius has the form
\be
{\bar r}_s=\frac{3(1-3s)\sin(\pi s)}{4 s \cos\left(\frac{3\pi}{2} s\right)}.
\ee

\subsection{Stability}

As we have just shown, the solutions \eqref{phi4} and \eqref{phi6} are obtained under specific boundary conditions, which suggest that they are stable. In spite of this, we now study stability of the solutions against spherically symmetric deformations. 
We study stability using that $\phi=\phi_s(r)+\epsilon\,\eta_s(r)$, with $\epsilon$ being very small real and constant parameter. We then expand the total energy in the form
\be
E_\epsilon =E_0+\epsilon E_1+\epsilon^2 E_2+\cdots
\ee
where $E_n, n=1,2,...$ is the contribution to the energy at order $n$ in $\epsilon$. For the model \eqref{p4}, $E_n$ goes up to $n=4$, and for the model \eqref{p6}, $E_n$ goes up to $n=6$. Of course, $E_0$ is the energy of the solution $\phi_s(r)$, and we can use the equation of motion to show that $E_1=0$. We have that the energy of the solution is
\be
E_0=2\pi \int_0^\infty r dr \left(\frac12 \left(\frac{d\phi_s}{dr}\right)^2+ \frac{1}{2r^2}P(\phi_s)\right).
\ee
For the model \eqref{p4} it gets to the form
\be \label{ene04}
E_0= \frac{8\pi}{3(1- s)}.
\ee
It reproduces the previous result \eqref{ene4}, as expected. Also, from $E_1=0$ we get that the zero mode is given by
\be
\eta_s(r)=A_s\frac{r^{2/(1-s)}}{(1+ r^{2/(1-s)})^2}.
\ee
Here, $A_s$ is a normalization constant, which depends on $s$. We can then show that $E_2=0$, $E_3=0$, and $E_4$ has the form
\be
E_4= \frac{3\pi}{35(1-s)}.
\ee
We see that it is positive, $E_4>0$, and this shows that the energy $E_0$ is a minimum of $E_\epsilon$ for the model \eqref{p4}, so the solution $\phi_s(r)$ which we obtained in \eqref{phi4} is stable against spherically symmetric fluctuations.
We follow the same steps for the model \eqref{p6}. We have that
\be
E_0=\frac{\pi^2 s}{2\sin(\pi s)},
\ee
as expected from \eqref{ene6}. Also, from $E_1=0$ we get that the zero mode is now given by
\be
\eta_s(r)=A_s\frac{r^{2/(1-s)}}{(1+r^{2/(1-s)})^{3/2}}.
\ee
We can show that $E_2=0$ and
\be
E_3=\frac{15\pi^ 2}{64(1-s)}.
\ee
We see that it is positive, $E_3>0$, and this shows that the spherically symmetric solution \eqref{phi6} is stable against spherically symmetric fluctuations.

The spherically symmetric solutions that we have found may also be disturbed by non spherically symmetric fluctuations. This would appear from the presence of defects and/or impurities in the magnetic material, and also from external fields. Such perturbations would break the spherical symmetry which we consider to be present in the system, so we do not study them here. We emphasize that we are supposing that the magnetic material is homogeneous along the $z$ direction, to make it a planar system, and that it behaves along the planar directions, in a way such that the spherical symmetry is effective in the plane.
\begin{figure}[t]
\centerline{\includegraphics[scale=0.35]{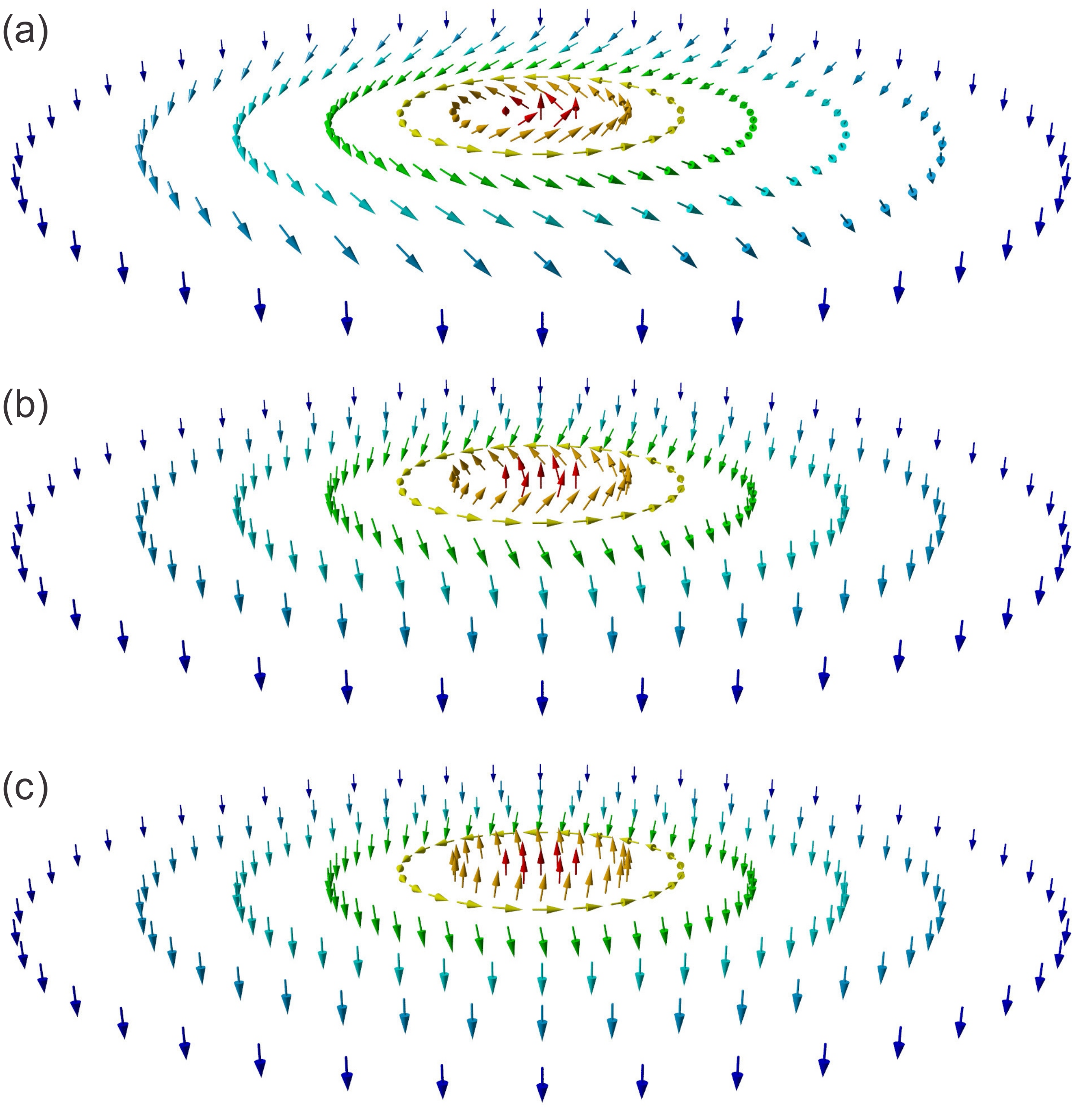}}
\caption{(Color online) The magnetization ${\bf M}$ for the model \eqref{p4}, depicted in (a), (b) and (c) for $s=0, 0.4$ and $0.8$, respectively.}
\label{fig3}
\end{figure}

\subsection{The planar solutions}
\label{planar}

With the assumptions discussed in the previous sections, we can now concentrate in the planar system, and search for an appropriate way to represent the magnetization. As we have stated in Sec.~II, we are dealing with helical excitations, with the magnetization
${\bf M}$ being a unit vector orthogonal to the radial direction. In the case of the model described by the polynomial \eqref{p4}, we have two distinct ground states, one with the magnetization vector pointing along the positive $z$ axis, and the other, along the negative $z$ axis. We can represent this situation with the magnetization \eqref{mm} with $\delta=0$, which gives
\be\label{mag4}
{\bf M}={\hat z}\sin\left(\frac{\pi}{2}\phi(r)\right)+{\hat\theta} \cos\left(\frac{\pi}{2}\phi(r)\right).
\ee
This shows that if $\phi$ is the uniform solution $\phi=-1$, we have the magnetization vector pointing downward in the ${\hat z}$ direction in the magnetic material. If the uniform solution is $\phi=1$, we have the magnetization vector pointing upward in the ${\hat z}$ direction in the magnetic material. These are the two possible ground states. However, if the field $\phi$ is represented by the solution \eqref{phi4}, we have another state, with energy given by \eqref{ene4}. If we take the magnetization vector as in \eqref{mag4}, with $\phi(r)$ given by \eqref{phi4}, we then have an array of vectors in the plane, which represents the helical magnetic excitation which is depicted in Fig.~\ref{fig3} for three distinct values of $s$. The magnetization vectors that we depict in Fig.~\ref{fig3} describe magnetic excitations supported by first model, investigated  above.

\begin{figure}[t]
\centerline{\includegraphics[scale=0.35]{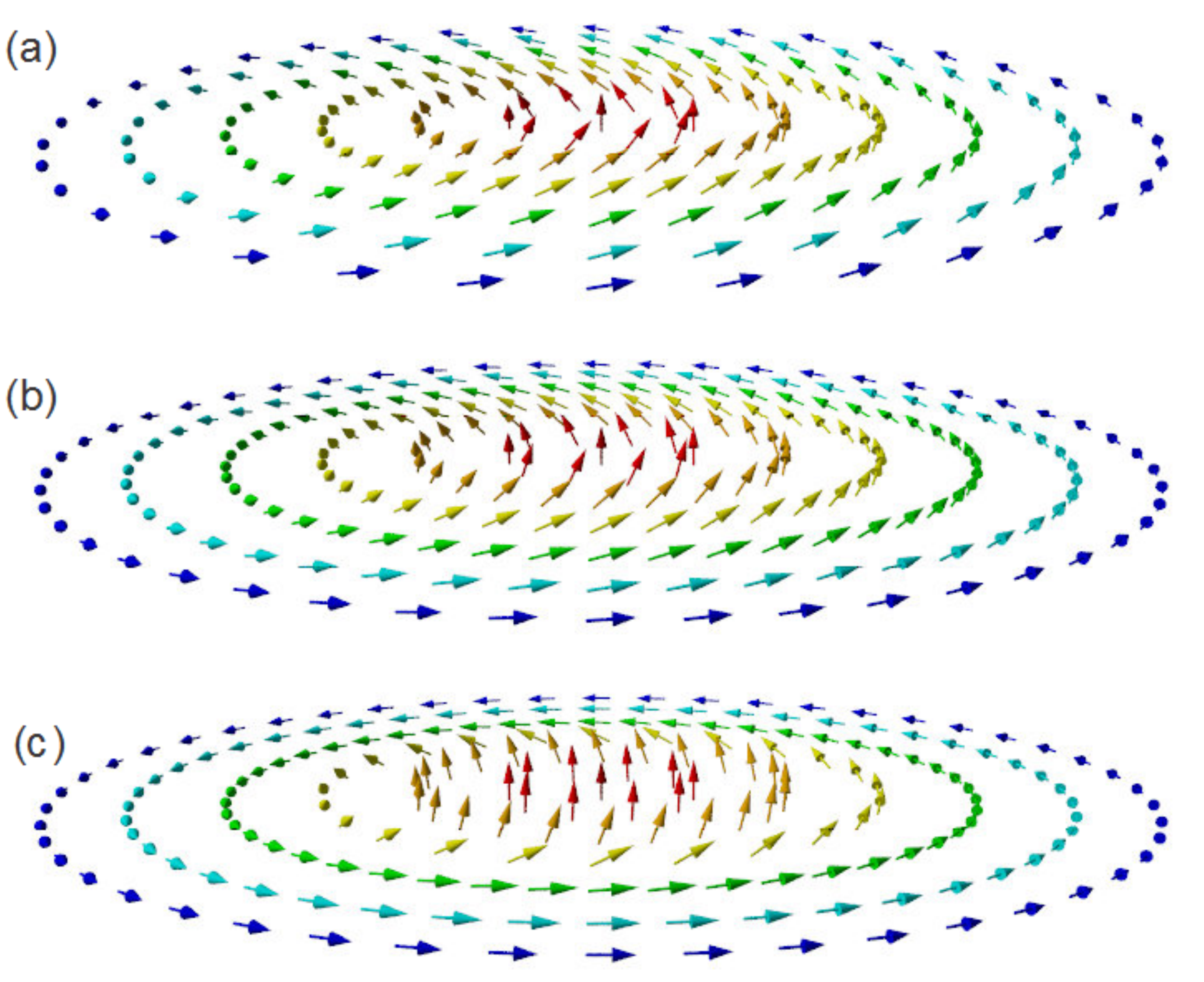}}
\caption{(Color online) The magnetization ${\bf M}$ for the model \eqref{p6}, depicted in (a), (b) and (c) for $s=0, 0.4$ and $0.8$, respectively.}
\label{fig4}
\end{figure}

We now concentrate on the second model \eqref{p6}. For $\phi$ being uniform field configuration, the model has three distinct ground states, one with $\phi=0$, and two with $\phi=\pm1$. Here we take the two ground states $\phi=\pm1$ represented by vectors pointing in the positive or negative sense, along the $\hat{\theta}$ direction. Also, we consider the uniform ground state $\phi=0$ represented by the magnetization vector along the ${\hat z}$ direction. With this in mind, we describe the magnetization \eqref{mm} with
$\delta=\pi/2$, in the form
\be\label{mm6}
{\bf M}={\hat z}\cos\left(\frac{\pi}{2}\phi(r)\right)+ {\hat \theta}\sin\left(\frac{\pi}{2}\phi(r)\right).
\ee
We then use the result in Eq.~\eqref{phi4} to show that the magnetization \eqref{mm6} describes an array of vectors in the plane, which represents the helical magnetic excitations that we depict in Fig.~\ref{fig4} for some values of $s$. 

\subsection{Skyrmion number}

Let us now concentrate on the skyrmion features of the solutions found above. 
The fact that we have found stable solutions with specific boundary conditions encourages us to investigate how to attain skyrmion number to them. We introduced the skyrmion number in Eq.~\eqref{sky}, which we now use to make it explicit. Since ${\bf M}={\bf M}(r)$, we change variables from $(x,y)$ to $(r,\theta)$ in Eq.~\eqref{sky} and use Eq.~\eqref{mm} to get the result
\be  
Q\!=\!-\frac12\sin\!\left(\frac{\pi}{2}\phi(\infty)\!+\!\delta\right)\!+\frac12\sin\!\left(\frac{\pi}{2}\phi(0)\!+\!\delta\right).
\ee
This is the expression we have to use to associate skyrmion number to the magnetic solutions that we found above. As expected, it only depends on the asymptotic behavior os the scalar field. Thus, since the asymptotic behavior of the scalar field is associated to the boundary conditions we used to solve the equation of motion, and since they are used to describe stable ground state solutions of the model under investigation, it appears that the above skyrmion number is a very natural choice in the current investigation. 

\begin{figure}[h!]
\centerline{\includegraphics[scale=1.2]{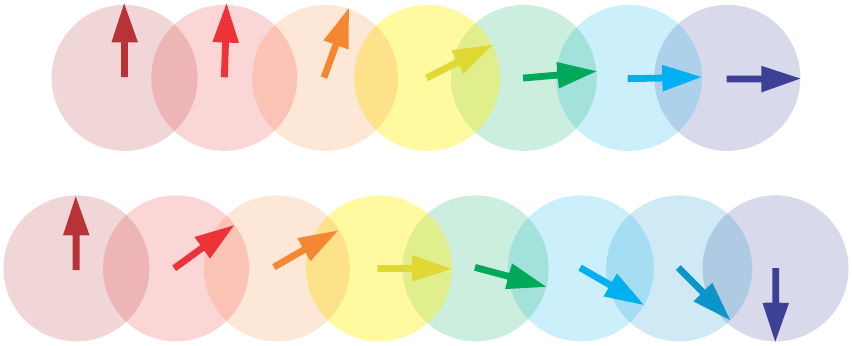}}
\caption{(Color online) Illustration of the magnetization vector ${\bf M}$ for the two models, depicted with $s=0$ for the model \eqref{p4} in the bottom panel, and with $s=0.8$ for the second model \eqref{p6} in the top panel.}
\label{fig5}
\end{figure}

To make the skyrmion number explicit, we now focus attention on the two models studied above. We consider the first model, where
$\phi(0)=1$, $\phi(\infty)=-1$, and $\delta=0$, to get $Q=1$. Thus, the magnetic excitation described in Fig.~\ref{fig5} has skyrmion number $Q=1$; the other magnetic excitation, with $\phi\to-\phi$ has skyrmion number $Q=-1$. For the second model, we have $\phi(0)=0$, $\phi(\infty)=1$, and $\delta=\pi/2$, and so we get $Q=1/2$; the other magnetic excitation, with $\phi\to-\phi$ has skyrmion number $Q=-1/2$. These solutions are then skyrmions, but they are different from each other. 

The solutions with $Q=\pm1/2$ are vortices \cite{vortex,vortex2}, and we see here that vortices ($Q=\pm1/2$) and skyrmions ($Q=\pm1$) appear under similar dynamical assumptions, controlled by the scalar field model \eqref{model}, but requiring distinct polynomial contributions, as noted from the two models \eqref{p4} and \eqref{p6}. This approach considers the dynamics of vortices and skyrmions qualitatively similar, described under the same framework, and is in accordance with the current understanding; see, e.g., Ref.~\cite{prb15}.

In order to highlight the difference between the two solutions, we note that we can associate a number of the chiral type to the localized excitations that we found in the second model, depicted in Fig. ~\ref{fig4}. We see from the solutions there depicted, that they have a chiral-like number $(+)$; the excitations obtained with the sign change $\phi\to-\phi$ would then have chiral-like number $(-)$. This is an interesting feature, which is not present in the first model, depicted in Fig.~\ref{fig3}. Thus, they may have different collective behavior, as we find if we want to construct a lattice of excitations, for instance. We can use the magnetic excitations of the first model to construct both the triangular and square lattices, but with the excitations of the second model, we will end up with frustration, if we want to construct the triangular lattice. This behavior is similar to the effect that appears in a triangular lattice of spins with antiferromagnetic interactions.

Before ending the work, it seems of interest to further highlight the helical profile of the magnetic excitations depicted in Figs.~\ref{fig3} and \ref{fig4}. We illustrate this in Fig.~\ref{fig5}, where we show how the magnetization rotates in the plane orthogonal to the radial direction, as a function of $r$. We depict the magnetization vector $\bf M$ with similar colors and the same values of $r$ that we have used to draw both Figs.~\ref{fig3} and \ref{fig4}. As it is clear, the magnetization has constant modulus, but it rotates in the $({\hat z},\hat{\theta})$ plane as $r$ increases to higher and higher values.
{\\}

{\section{Ending comments}

In this work we studied the presence of localized structures of the skyrmion type in planar magnetic materials. We described the magnetization vector ${\bf M}$ in terms of a single degree of freedom, in the case of helical excitations, in which the magnetization has constant modulus but is allowed to rotate in the $({\hat z}, {\hat\theta})$ plane, orthogonal to the radial direction.  The topological excitations that we found are skyrmions, and we could associate a skyrmion number to them. 

We used the scalar field $\phi$, which is described by the model of Ref.~\cite{bmm} in the case of two spatial dimensions, to control the magnetization ${\bf M}$, in the plane orthogonal to the radial direction. The assumption that the magnetization is orthogonal to the radial direction is appropriate to describe helical excitations in magnetic metals, as suggested before in \cite{s2,s3} and identified experimentally for instance in \cite{e1}.

All the results obtained in the current work are constructed analytically, and they seem to map magnetic excitations very appropriately. For the  second model, defined from \eqref{p6}, we have identified magnetic excitations that present a chiral feature, being different from the magnetic excitations that appear in the first model, defined from \eqref{p4}.
The study encourages us to go further and investigate other models, described with different polynomials $P(\phi)$, contributing to generate magnetic excitations with distinct profile. Also, we can study models described by two or more real scalar fields, leading to a richer distribution of minima, which could be of good use to describe more general scenarios; see, e.g., Ref.~\cite{BB,BB2} for some studies of kinks with two real scalar fields in one spatial dimension, that can be seen as background for the investigation in two spatial dimensions. 

We can also consider the case where the size of the magnetization vector is allowed to vary along the radial direction. Another issue of current interest concerns the formation of lattices of magnetic solitons. As we have commented at the end of the previous section, an interesting situation occurs if we think of constructing a triangular lattice of excitations which appear in the second model, since the triangular configuration would conflict with the boundary conditions required to support each excitation in the lattice. These and other related issues are presently under consideration, and will be reported elsewhere.

This work is partially supported by the Brazilian agency CNPq. DB thanks support from grants CNPq:455931/2014-3 and CNPq:06614/2014-6, and MMD and EIBR acknowledge support from grants CNPq:23079.014992/2015-39 and CNPq:160019/2013-3, respectively.


\end{document}